\documentclass[review]{elsarticle}

\journal{Nuclear Instruments and Methods A}

\begin{document}

\begin{frontmatter}

\title{X-ray tracing using Geant4}

\author{E.~J.~Buis}
\author{G.~Vacanti\corref{cor1}}
\ead{g.vacanti@cosine.nl}
\ead[url]{cosine.nl}
\address{cosine Science \& Computing BV, Niels Bohrweg 11, 2333 CA Leiden, The Netherlands}
\cortext[cor1]{Corresponding author. Tel. +31 71 5284962; fax. +31 71
  5284963.}

      \begin{abstract}   
	We describe an extension to the Geant4 software package that
        allows it to be used as a general purpose X-ray tracing
        package. We demonstrate the use of our extension by building a
        model of the X-ray optics of the X-ray observatory XMM-Newton,
        calculating its effective area, and comparing the results with
        the published calibration curves.
      \end{abstract}

\end{frontmatter}

    
\section{Introduction}
\label{sec:introduction}

The Geant4 simulation toolkit is a software library for high-energy
physics developed by an international collaboration under the lead of
CERN~\cite{G4}. Aiming at supporting the simulation and analysis of
data collected by the Large Hadron Collider, Geant4 has found a keen
following in many other areas where the ability to model complex
geometries and the interaction of particles and radiation with matter
plays a prominent role. The open nature of the toolkit, and the large
libraries of geometrical models and physical processes that are
available make it easy to use and extend.

The importance of detailed ray-tracing simulations in the development,
construction, testing, and operation of X-ray optics has long been
recognized. First described by Wolter in
1952~\cite{wolter52:_glanc_incid_mirror_system_as}, the X-ray optics
that would later be given his name were quickly adopted as the
preferred configuration for X-ray imaging
systems~\cite{1960JGR....65..773G}. In 1972, Chase and
Van Speybroeck~\cite{1972ApOpt..11..440V} carried out the first
numerical evaluation of the optical properties of Wolter
optics. Since then several authors have investigated X-ray imaging
systems (see for instance \cite{saha03:_equal_curvat} and references
therein).

To a large extent X-ray imaging systems can be treated as any other
optical system, and their properties in terms of aberrations and
focusing capabilities can be examined with the standard tools of
optics.

However, when detailed performance predictions of the effective area
and the point spread function are required, specific ray-tracing tools
are needed in order to take into account the change of the
reflectivity with the angle of incidence, and the scattering caused by
the details of the surface finish. To the best of our knowledge, and
certainly within the domain of X-ray optics for astrophysical
missions, this problem has always been solved with ad hoc
tools built to address issues related to the particular design under
consideration.

We have developed a set of extensions for the Geant4 toolkit that
allows one to trace X-ray optics of arbitrary complexity, as long as
the geometrical model of the system can be described by the geometry
library available in Geant4. The core component of these extensions
consists of a Geant4-compatible implementation of the reflection of
X-ray photons on a surface. Additionally, by making use of the
available Geant4 functionality, we have created a framework that
allows one to introduce a microscopic description of the properties of
the surfaces of the optics. Through this framework it becomes possible
to model the effects that these microscopic properties have on the
scattering of the photons.

Besides introducing the notion of a generic tracer for X-ray optics,
this development has the potential to ease the interaction of
scientists and engineers during the study phase of X-ray space
missions. Geant4-based models are in fact already widely used to study
the effect of cosmic radiation on the spacecraft structures and
instruments. By making use of the same tools in order to model both
the spacecraft and the telescope, the study of a mission can be
carried out on the basis of a shared understanding of its mechanical
and optical properties.

This article is organized as follows. In
\S\,\ref{sec:reflection-x-rays} we briefly describe the physics of
X-ray reflection and scattering at grazing angles. In
\S\,\ref{sec:x-ray-oriented} the X-ray oriented extensions to the
Geant4 toolkit that we have implemented are described. We proceed then
to describe a sample application of our extensions in
\S\,\ref{sec:application-example}. Finally, in
\S\,\ref{sec:concl-future-extens} we draw some conclusions and make
some final remarks.

\section{The Reflection of X-Rays}
\label{sec:reflection-x-rays}

\begin{figure}[tpb]
  \begin{center}
    \includegraphics[width=7cm]{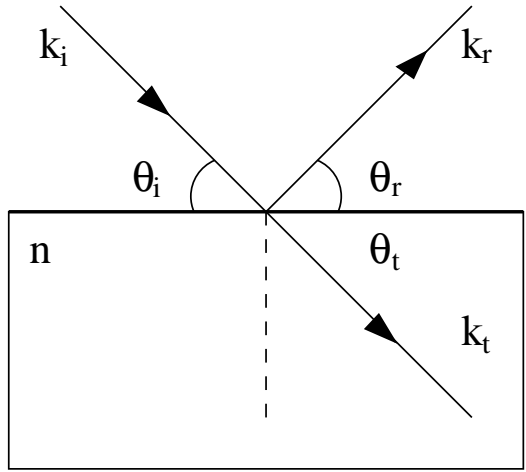}
    \caption{\label{fig:fresnel} Schematic representation of the
      reflection and transmission at the boundary between vacuum and
      material.}
    \end{center}
\end{figure}

The reflection of X-rays at the boundary between vacuum and a medium
(see Fig.~\ref{fig:fresnel})
is described by the standard Fresnel equations (see for instance
\cite{bohrwolf}). Here, in keeping with the usage in the field, we use
energy instead of wavelength.

For a photon with grazing incidence angle $\theta_i$ and energy
E, the reflection coefficient $R$ and the transmission coefficient $T$
(both for the orthogonal and parallel polarizations) are:
\begin{eqnarray}
  \label{eqn:fresnel}
  R(\theta_i, E)  &=& \left| \frac{\sin \theta_i 
    - n(E)\sin \theta_t}
  {\sin \theta_i + n(E)\sin \theta_t}\right|^2
  \nonumber \\
  T(\theta_i, E)  &=& \left| \frac{2\sin \theta_i}
  {\sin \theta_i + n(E)\sin \theta_t}\right|^2\,,
\end{eqnarray}
where $n(E)$ is the medium's index of refraction. For different
materials $n(E)$ can be obtained from various sources. In practice, the
reflectivity is calculated using experimentally determined optical
constants that are valid in the X-ray regime. However, when these
optical constants are not available because of the lack of
measurements (in the case of non standard materials) or because the photon
energy is in the gamma-ray regime, where only a few optical constants
are known, one can construct the optical constants using the atomic
form factors (e.g.~\cite{1993ADNDT..54..181H}).

The reflection coefficient given in Eq.(\ref{eqn:fresnel}) is
valid in the case of an ideal, perfectly smooth surface. For a real
surface, some fraction of the reflected photons will be scattered away from
the specular direction. Under assumptions generally true for X-ray
optics, it is customary to characterize a surface by its
micro-roughness $\sigma$, and decrease the reflectivity by the factor

\begin{eqnarray}
  \exp\left(-\left(\frac{4\pi\sigma\sin\theta_i}{\lambda}\right)^2\right)\,.
  \label{eq:1}
\end{eqnarray}

In the context of astrophysical applications of X-ray optics,
Eq.(\ref{eq:1}) is sufficient to perform calculations related to
the effective area of a particular optical design. More complex
calculations aiming at the prediction of the point spread function
must use a more extensive theory (for instance
\cite{beckman_scattering_1963}, \cite{croce76}), whose details are
affected by the type of surface data available (e.g., surface
interferograms, profilometry measurements, atomic force telescope
data). Our software implementation provides a generic interface for
the application of any scattering model, as is described in the
following section.

\section{X-Ray-Oriented Extensions of the Geant4 Toolkit}
\label{sec:x-ray-oriented}

The extension of the Geant4 toolkit with the physics of grazing angle
scattering is realized through the implementation of three classes:

\begin{description}
\item[G4XrayRefractionIndex] 

  This is an auxiliary class that manages the refraction index data
  for a particular material.

\item[G4XraySurface]

  As mentioned in the previous section, the details of the interaction
  of X-rays on a surface can be described by scattering models of
  various complexity, and are also driven by the type of data describing the
  microscopic surface details. It is therefore not surprising that
  even within the same model different surfaces may have different
  X-ray behaviours: for instance, an X-ray mirror will have one
  reflecting side, and a back side treated so as to absorb or scatter
  away from the main beam all photons that might fall on it. Also,
  when assembling an optical system for which disparate surface data
  types are available, it may be desirable to be able to assign
  different scattering models to different surfaces. The class
  G4XraySurface provides a generic interface to do exactly this. It
  can be used with the standard Geant4 mechanisms to define either a
  logical boundary surface, or a logical skin surface. We have
  written sample classes that implement Fresnel reflection as
  described in the previous section, and that perturb the surface
  normal at the interaction point according to a Gaussian
  distribution. In general, G4XraySurface can be used to implement any
  scattering model appropriate for the particular situation
  considered.

\item[G4XrayGrazingAngleScattering]

  This class models a new Geant4 boundary process, and is applicable
  to any photon (the Geant4 particle gamma: no new particle needs to
  be defined). The process can compete with any other Geant4 processes
  that might be defined in the simulation (for instance photo-electric
  effect, or Compton scattering). When the Geant4 tracking system
  detects that a photon is crossing the boundary between two volumes
  (normally from vacuum to a material), this process is invoked. In
  turn, the process queries the surface being crossed to see if it is
  an X-ray surface: if this is the case, the process delegates the
  actual interaction details to the surface. The scattered/reflected
  photon is handed back to the process who decides its fate---absorbed
  or transmitted, reflected, or propagated further---, and eventually
  hands the control back to the Geant4 tracking manager. The process
  can be activated so as to log the details of every X-ray
  interaction: the coordinates of the interaction point, the volume on
  which the interaction has taken place, the (possibly modified) local
  surface normal, incoming and outgoing directions of the photon, and
  the reflectivity used. This information, that is commonly used in
  the field to understand the details of the geometrical model and how
  certain features on the focal plane are formed, can be accessed
  using the standard track information facilities of Geant4.
\end{description}

\begin{figure}[tpb]
  \setlength{\unitlength}{1cm}
  \begin{minipage}[b]{0.5\textwidth}
    \includegraphics[width=\textwidth]{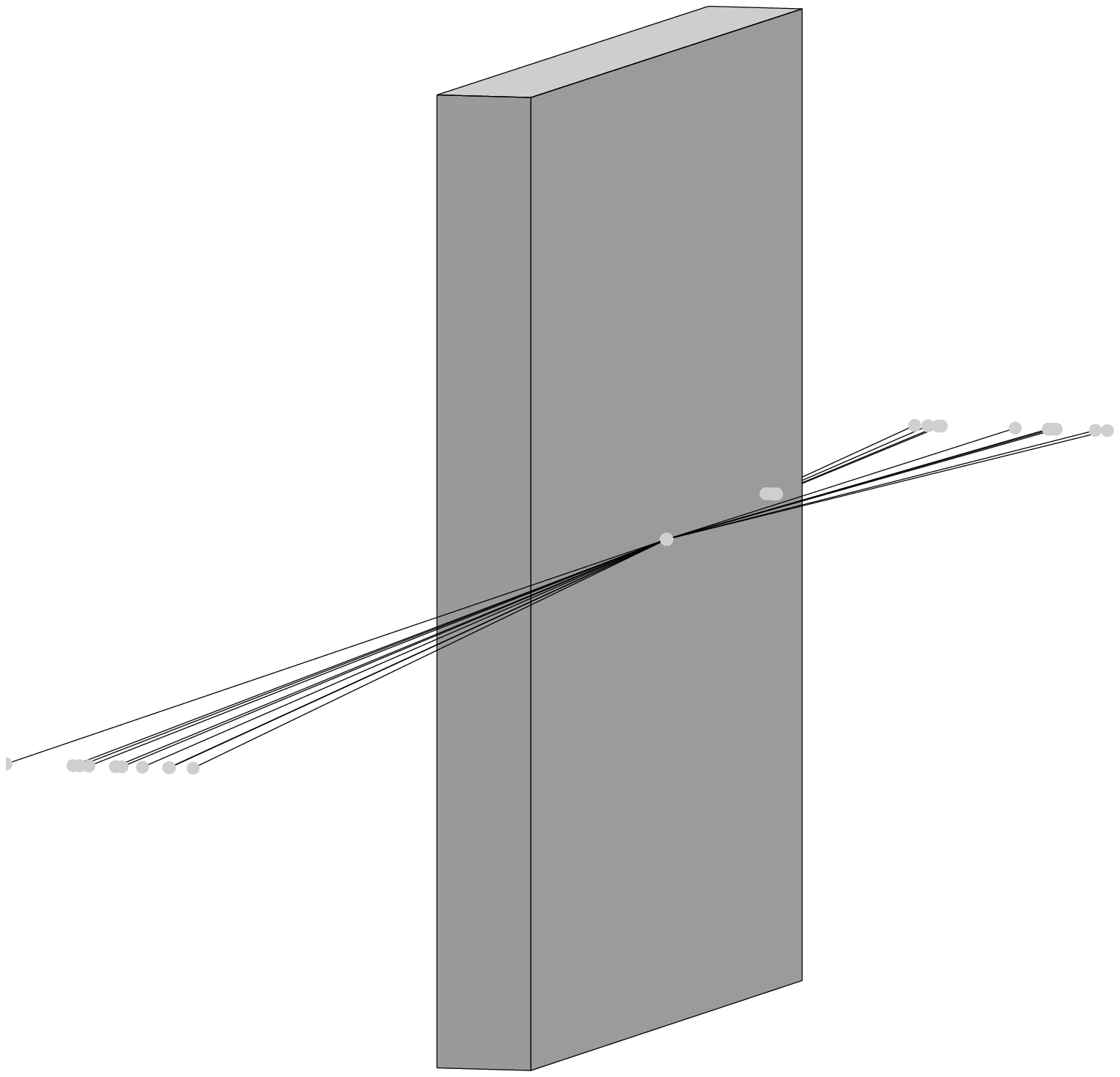}
  \end{minipage}\hfill
  \begin{minipage}[b]{0.5\textwidth}
    \includegraphics[width=\textwidth]{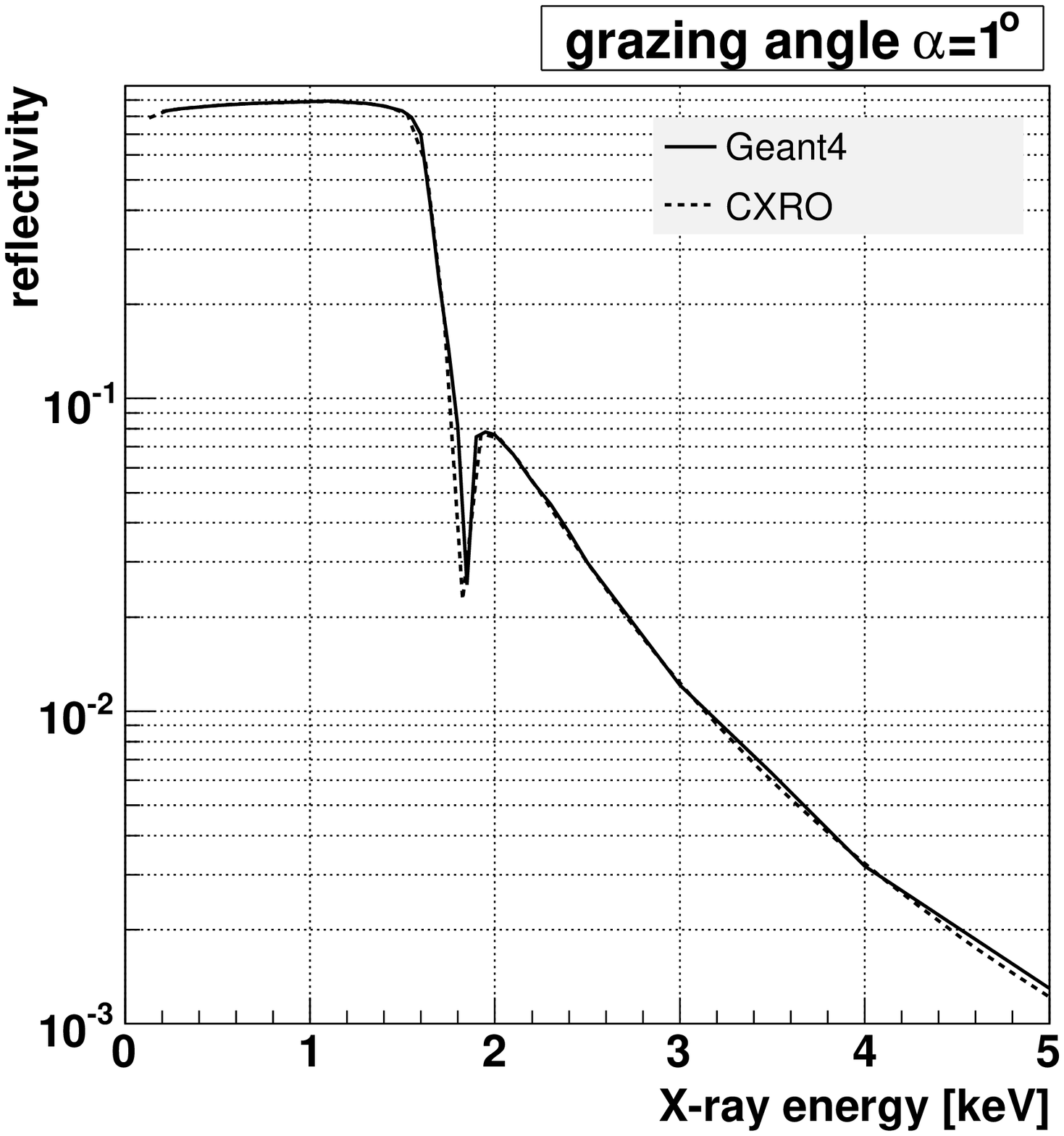}
  \end{minipage}
  \begin{minipage}[t]{0.5\textwidth}
    \begin{center}
      (a)
    \end{center}
  \end{minipage}
  \hfill
  \begin{minipage}[t]{0.5\textwidth}
    \begin{center}
      (b)
    \end{center}
  \end{minipage}
  \begin{minipage}[b]{\textwidth}
    \caption{\label{fig:si_refl} (a) Simulated reflectometry
      setup. (b) The reflectivity determined from the simulation is
      compared with data retrieved from the Center for X-ray
      Optics~\cite{cxro}.}
 \end{minipage}
\end{figure}

Using the extension a simple reflectometry set up can be constructed
consisting of a single silicon plate as shown in
Fig.~\ref{fig:si_refl}(a). The reflectivity for a fixed angle of
incidence can then be determined, and compared with the data obtained
using the tools made available by the Center for X-Ray
Optics~\cite{cxro}, a standard reference in the field. In
Fig.~\ref{fig:si_refl}(b) the two data sets are shown to be in very
good agreement.

\section{Application Example}
\label{sec:application-example}

\begin{figure}[tpb]
  \setlength{\unitlength}{1cm}
  \begin{minipage}[b]{0.5\textwidth}
    \begin{center}
      \includegraphics[width=0.9\textwidth,angle=90]{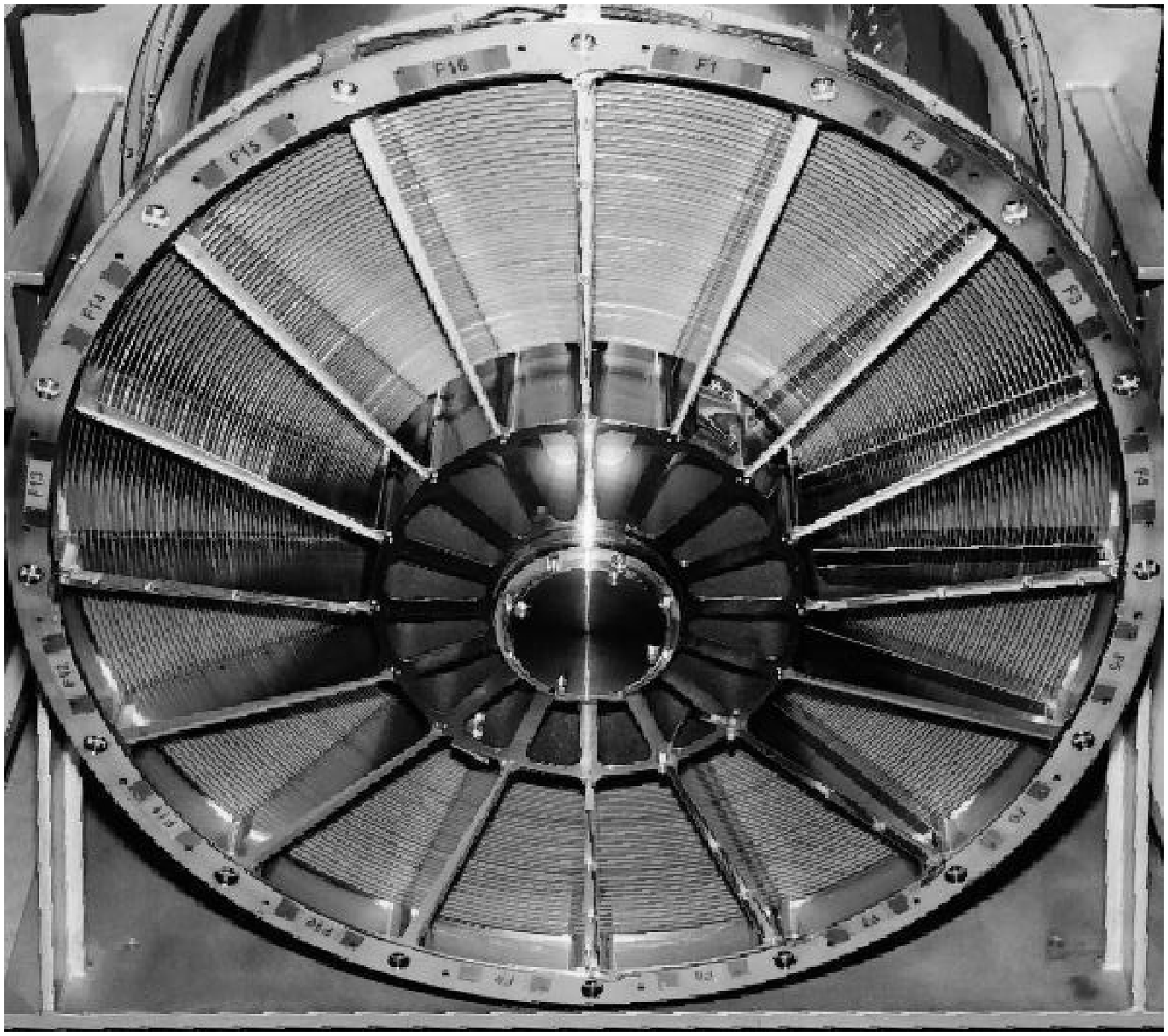}
    \end{center}
  \end{minipage}\hfill
  \begin{minipage}[b]{0.5\textwidth}
    \begin{center}
    \includegraphics[width=\textwidth]{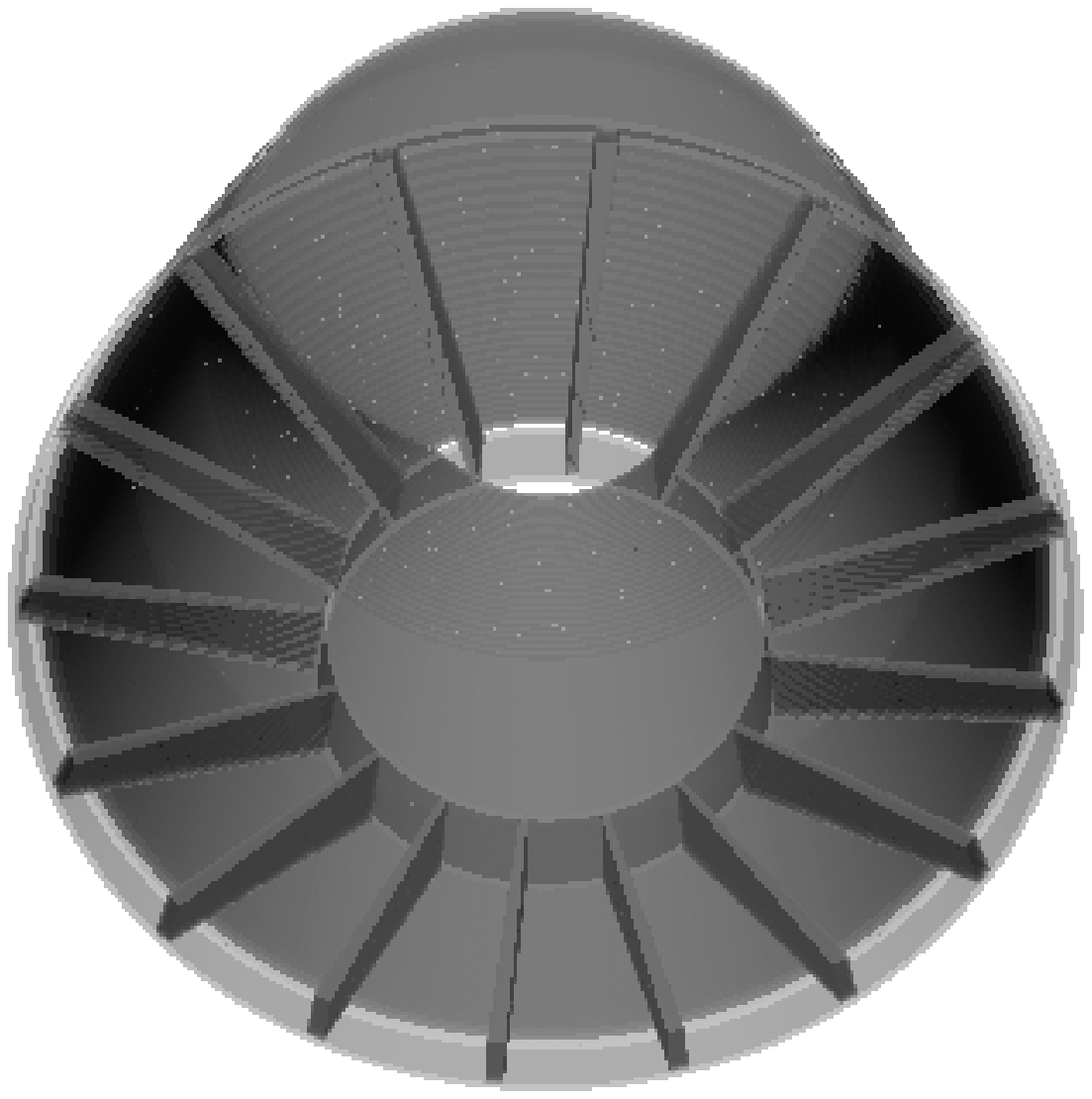}
    \end{center}
  \end{minipage}
  \begin{minipage}[t]{0.5\textwidth}
    \begin{center}
      (a)
    \end{center}
  \end{minipage}
  \hfill
  \begin{minipage}[t]{0.5\textwidth}
    \begin{center}
      (b)
    \end{center}
  \end{minipage}
  \begin{minipage}[b]{\textwidth}
    \caption{\label{fig:xmm_geom} (a) Photograph of the front view of
      the XMM-Newton telescope (courtesy of the European Space
      Agency). (b) The Geant4 model used for the comparisons.}
 \end{minipage}
\end{figure}

As an example we have modeled the X-ray optical system of the
XMM-Newton telescope. Each of the three mirrors aboard XMM-Newton
comprises 58 concentric mirror shells in a Wolter-I configuration,
where each mirror shell consists of polished, confocal parabolic and
hyperbolic surfaces. The parabolic and hyperbolic solids that are
included in the Geant4 package are not directly useful to construct a
Wolter-I system, because they are filled solids. We therefore
approximated them with many consecutive short conical shell segments
with the appropriate slope.  The front view of the telescope is shown
in Fig.~\ref{fig:xmm_geom}(a). In this photograph the nest of mirror
shells can clearly be seen together with the spider-like mechanical
support structure. The total aperture ranges from 150\,mm to
350\,mm. In Fig.~\ref{fig:xmm_geom}(b) the same geometry is shown as
implemented in our model.  The mirror shells are coated with a gold
layer and we assume a surface roughness of 0.5\,nm root-mean-square.

\begin{figure}[tpb]
  \setlength{\unitlength}{1cm}
  \begin{minipage}[b]{0.5\textwidth}
    \includegraphics[width=7cm]{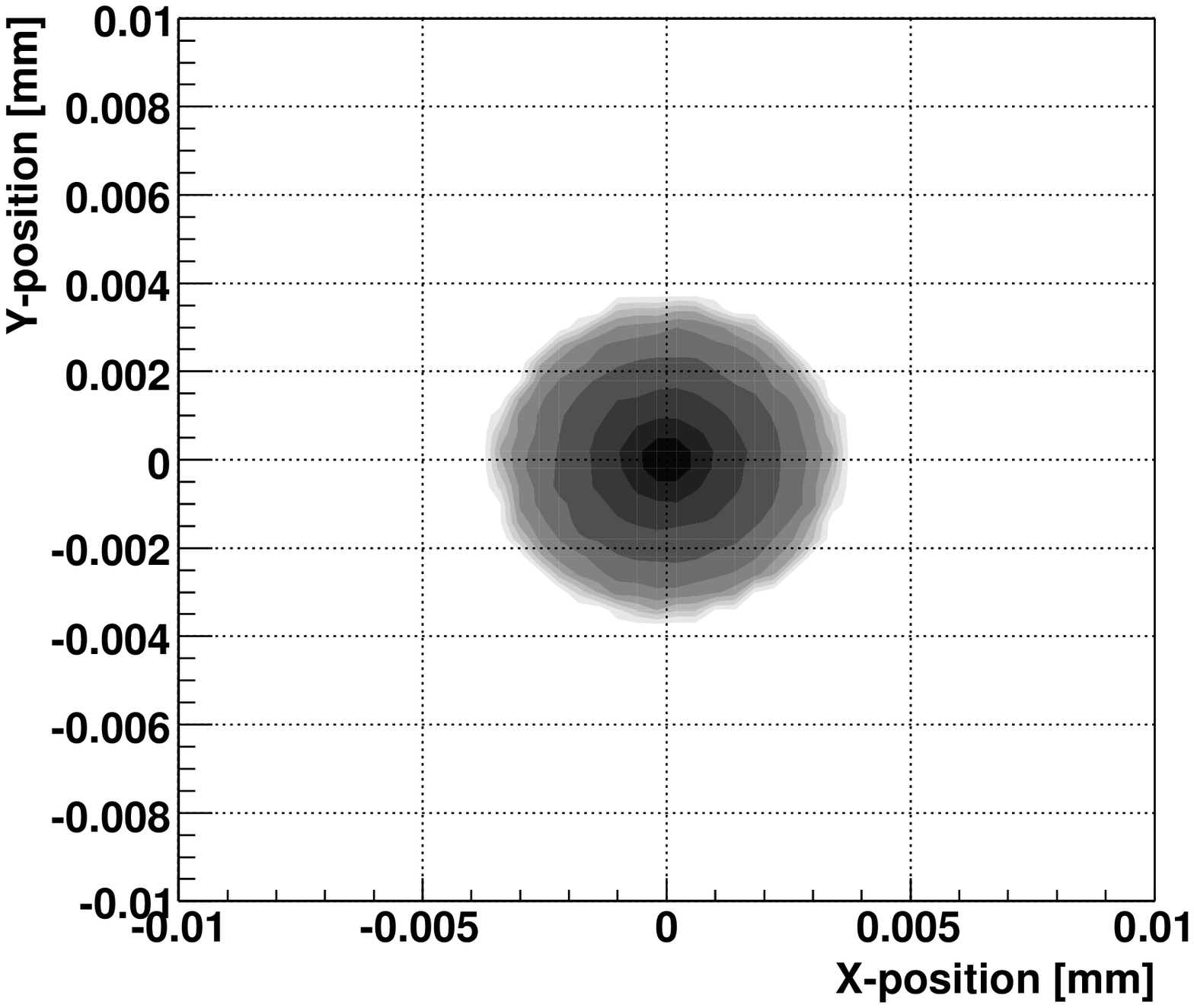}
  \end{minipage}\hfill
  \begin{minipage}[b]{0.5\textwidth}
    \includegraphics[width=7cm]{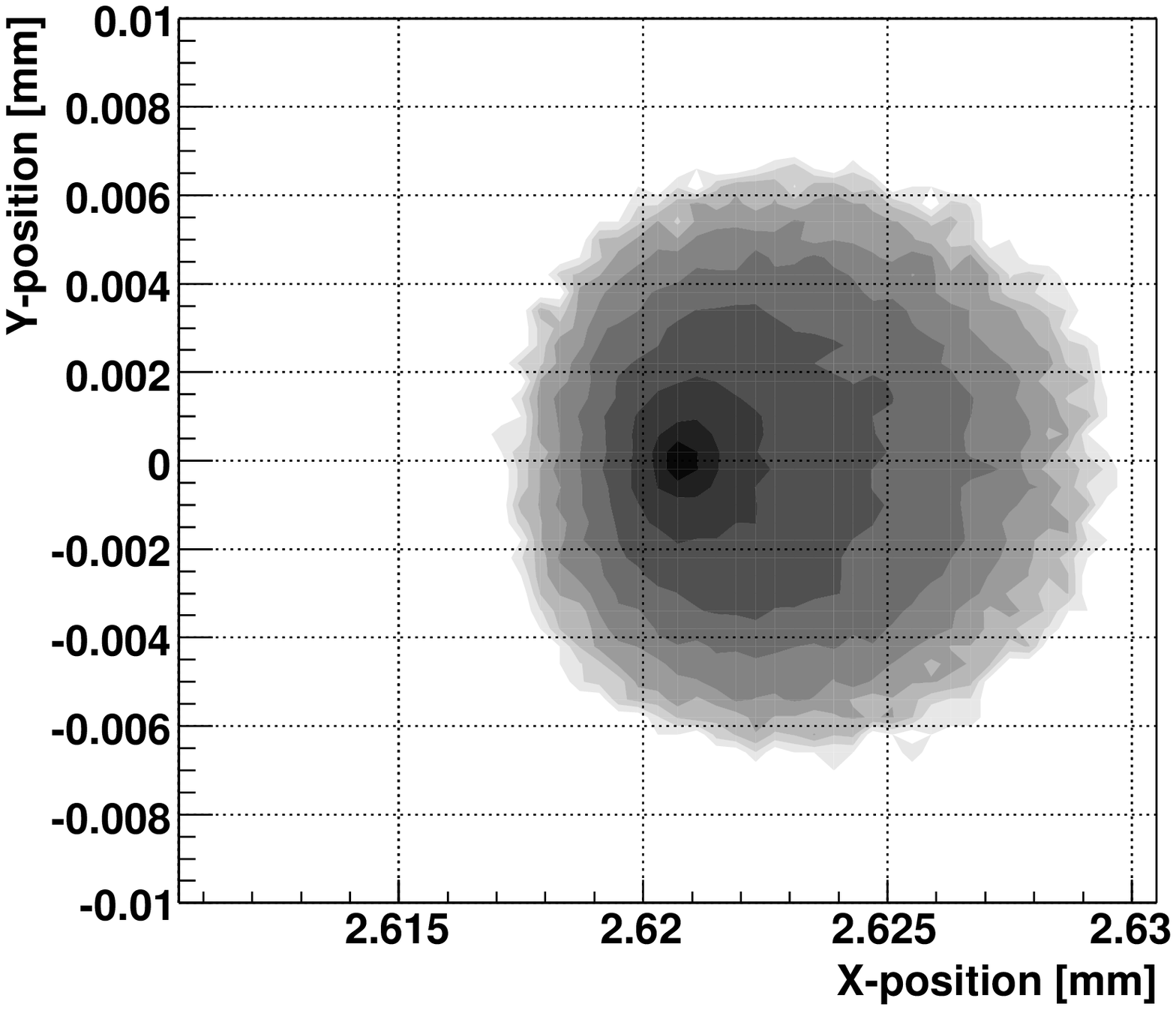}
  \end{minipage}
  \begin{minipage}[t]{0.5\textwidth}
    \begin{center}
      (a)
    \end{center}
  \end{minipage}
  \hfill
  \begin{minipage}[t]{0.5\textwidth}
    \begin{center}
      (b)
    \end{center}
  \end{minipage}
  \begin{minipage}[b]{\textwidth}
    \caption{\label{fig:xmm_focal_spot} (a) Focal spot of the XMM-Newton
    telescope for an on-axis point source. (b) Focal spot for an
    off-axis (0.02\,degrees) point source.}
 \end{minipage}
\end{figure}

In Fig.~\ref{fig:xmm_focal_spot} we show the focal spot for an on-axis
X-ray beam in comparison to the focal spot for an off-axis beam.  The
spot for the off-axis beam lies away from the center of the image
plane as expected. For an off-axis angle of $2\cdot10^{-2}$\,degrees
and a focal length of 7500\,mm, the x-coordinate is expected to be
2.618\,mm, in agreement with the simulated result of 2.622\,mm.  Note
further that the shape of the spot becomes asymmetric due to coma, as
expected.

\begin{figure}[tpb]
  \begin{center}
    \includegraphics[width=\textwidth]{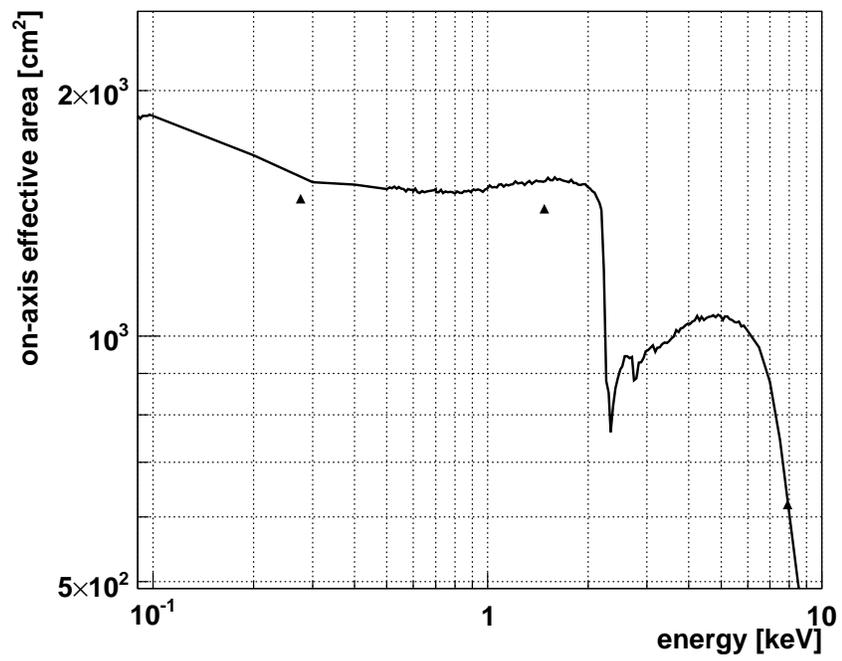}
    \caption{\label{fig:xmm_Aeff} On-axis effective area of the
      XMM-Newton telescope. The solid line shows the results of our
      simulation. The three data points (triangles) are taken
      from~\cite{Xmm_at_Panter}. The difference between measured and
      simulated effective area is less than 5\%.}
    \end{center}
\end{figure}

By Monte Carlo simulation we have determined the effective area of the
XMM-Newton optics, by taking the blocking effect due to the
mechanical support structure into account. We have not taken into
account any other effects, like dust contamination, or the exact form
of the mirror shells. In Fig.~\ref{fig:xmm_Aeff} we compare our
results with the ground calibration measurements reported in
\cite{Xmm_at_Panter}: our results are in very good agreement with the
data, differing no more than 5\% from them.

\section{Conclusions}
\label{sec:concl-future-extens}
We have created an extension of the Geant4 toolkit that makes it
suitable as a platform with which to realize generic X-ray
tracers. Besides modelling the grazing angle reflection of X-rays on
surfaces, our extension can be used to model the effects of surface
finish on the scattering of radiation.

We have demonstrated the new capabilities of the toolkit by modelling
one of the Wolter-I mirrors used on the European Space Agency's X-ray
observatory XMM-Newton. Without having access to the detailed
metrological data that were used during the calibration of XMM-Newton,
we are able to reproduce the measured effective area to about 5\%. We
have also shown that the main properties of the point spread function
are well reproduced.  Being aware of the efforts that went into
accurately modelling the response of the XMM-Newton mirrors
\cite{beijersbergen08}, we consider our result as a demonstration of
the potential our work has to ease the design and calibration of future
X-ray optics.

With this extension we have turned Geant4 into a versatile modeler for
X-ray optics.

\section*{Acknowledgments}

This work was carried out under contract to the European Space Agency.


\newpage

\listoffigures{}

\end{document}